# The internal structure of the Uranian satellites: a preliminary note

Vladan Celebonovic

Institute of Physics, Pregrevica 118,11080 Zemun-Beograd, Yugoslavia

vladan@ phy. bg. ac. yu
vcelebonovic@ sezampro. yu

Abstract: Preliminary results of a theoretical determination of the basic parameters of the internal structure of the five Uranian satellites are presented.

Four satellites of Uranus were discovered by Sir William Herschel in 1781.;the fifth was found by Kuiper in 1949. Reliable values of their masses and radii were determined only recently, after the passage of Voyager 2 through the Uranian system. By combining results of the Radio Science Team, with the star-satellite imaging data and with 8 years of ground based observations, it became possible to determine the radii and densities of the satellites with relative errors in some cases as low as 4 % ( Anderson et al.,1987 ). The best pre-Voyager results (Dermott and Nicholson,1986) had relative errors going up to 45 %.Such a situation has until recently severely hampered ( Prentice 1986;Anderson et al.,1987 ) any theoretical considerations of the internal structure and chemical composition of these bodies.

The purpose of this note is to present some preliminary results of a theoretical determination of the basic parameters of the internal structure of the five Uranian satellites. The calculations were performed within a particular semiclassical theory of dense matter (Savic and Kasanin,1962 / 65 ). Physically speaking, the main idea of this theory is that high pressure can cause excitation and ionisation of atoms and molecules; this process can be considered in detail quantum mechanically. Various examples of applications of astrophysical applications of this theory have already been published ( such as Savic,1981;Savic and Teleki,1986;Celebonovic,1988b and references given therein). A comparison of predictions of this theory with high pressure experiments in diamond anvil cells has recently given promising results (Savic and Urosevic,1987 ; note that eq.(i ) of this paper must be divided by two).

The input data for the calculation (i.e the masses and radii of the satellites) were taken from (Anderson et al.,1987).Starting from these data,and using the approach proposed by Savic and Kasanin,the following parameters of the satellites were derived:

TABLE I

| satellite | A (amu) | V (cm$^3$) | p$^*$ (kbar) | a (10$^{-3}$ AU) | $\rho$ (g cm$^{-3}$) |
|---|---|---|---|---|---|
| Titania | 32±4 | 19±2 | 101±3 | 2.9303 | 1.685±0.008 |
| Oberon | 32±1 | 20±1 | 97±4 | 3.9178 | 1.635±0.006 |
| Umbriel | 44±6 | 28±8 | 60±10 | 1.7860 | 1.58±0.23 |
| Ariel | 43±6 | 28±8 | 62±10 | 1.2820 | 1.55±0.22 |
| Miranda | 38±10 | 30±16 | 55±15 | 0.872 | 1.25±0.33 |

The satellites are arranged in order of diminishing radii. A and V denote, respectively, the mean atomic mass and the molar volume under standard conditions of the material that a satellite is made of; the central pressure is denoted by $p^*$, and a is the semiaxis major of the satellite's planetocentric orbit (Allen,1973).

Several qualitative conclusions can be drawn from data presented in Table I.

One can, for example, compare the values of A derived in this note, with those obtained earlier for various other bodies in the planetary system (Celebonovic,1988b and references given there ). It turns out that by their values of A, the satellites of Uranus are situated between the Earth and Mars. However, their observed densities are 2-5 times lower than for the two planets. This discrepancy can be interpreted as a result of the presence of a large proportion of ices ( $H_2O; NH_3; CH_4....$) in these satellites.

Another interesting result is the existence of gradients of A and r. Assuming that all the five satellites formed in the vicinity of Uranus, their present values of A and r reflect the distribution of ices and heavier chemical compounds in the Uranian system ate the time of its formation. Details of these distribution functions are heavily model dependent.

For example, in the so-called modern laplacian theory (Prentice,1986,and references given there),the satellites condensed from a system of orbiting gas rings that were shed by a gravitationally contracting parent envelope. This envelope disposes of its excess spin by sheding mass at the equator in discrete amounts and at discrete orbital radii. The temperature of the gas rings at the moment of detachement from the parent cloud varies with the sheding radius , and this could be the physical basis for the explanation of the compositional gradients in the Uranian satellite system.

In the cosmogonical model proposed by Alfven and Arrhenius (Alfven,1986;Alfven and Arrhenius,1985 and numerous preceeding publications),the formation of planets and/or satellites is explained by by invoking the so - called critical velocity, achieved by material falling towards a central body.
This seems to account for the band structure of the planetary and most satellite systems. If the primordial Uranian proto-satellite cloud consisted of a mixture of ices and rock, the critical velocity phenomenon could have easily led to the formation of two compositionally different groups of satellites.

Note ( added December 13,1998) : This paper was originally published in Bull. Obs. Astron Belgrade, **140** ,47 (1989). For an account of later work within the theory of Savic and Kasanin, see astro-ph/9803213 .

References

Alfven, H. and Arrhenius, G.: 1985,preprint TRITA-EPP-85-04.
Alfven, H.: 1986,IEEE Trans. on Plasma Sci.,**PS14**, 629.
Allen, C.W.: 1973,Astrophysical Quantities, Univ. of London Press, London.
Anderson, J.D., Campbell, J.K. ,Jacobson, R.A. et.al.:1987,J.Geophys. Res.,**92A**, 14877.
Celebonovic, V.: 1988b,Earth,Moon and Planets,**42**, 297.
Dermott, S.F. and Nicholson,P.D.: 1986,Nature,**319** , 115.
Prentice,A.J.R.: 1986,Phys.Lett.,**114A** , 211.
Savic,P. and Kasanin,R.: 1962/65,The Behaviour of Materials Under High Pressure I-IV, Ed. by SANU, Beograd.
Savic,P.: 1981,Adv.Space Res.,**1**,131.
Savic,P. and Teleki,G.: 1986,Earth,Moon and Planets,**36**, 139.
Savic,P. and Urosevic,V.: 1987,Chem.Phys.Lett.,**135**,393.